# A Blueprint of IR Evaluation Integrating Task and User Characteristics: Test Collection and Evaluation Metrics

Kal Jarvelin & Eero Sormunen, Tampere University

Faculty of Information Technology and Communication Sciences | Communication Sciences, Tampere University, FI-33014 Tampere University, Finland

ORCID: https://orcid.org/0000-0001-7655-8930 (Jarvelin)

**Abstract:** Traditional search result evaluation metrics in information retrieval, such as MAP and NDCG, naively focus on topical relevance between a document and search topic and assume this relationship as mono-dimensional and often as binary. They neglect document content overlap and naively assume gains piling up as the searcher examines the ranked list at greater length. The searcher remains looming on backstage. This blueprint proposes a novel test collection and search result evaluation metric based on multidimensional, non-binary relevance assessments, explicit modelling of document overlaps and attributes affecting document usability beyond relevance for the searcher. Document relevance to a search task in context is seen to consist of one or more content themes and null or more document usability attributes. A given document may to varying degrees be relevant to the content themes. Documents may also overlap to varying degrees regarding their content themes. Attributes such as document readability, trustworthiness, or language represent the entire document's usability in the search task context, for a given searcher and her/his motivating task, e.g., a school learning task or scientific research project. The proposed metric evaluates the quality of a ranked search result, taking into account the contribution of each successive document, with estimated overlap across themes, and usability based on its attributes. The number and kind of content themes depend on each search task and may therefore vary between search tasks of an evaluation study. The number and kind of attributes is fixed in the task context and likely stable between search tasks.

## 1. Introduction

Relevance is one basic concept of information science. It is intuitively understood but hard to define precisely. Nevertheless, it is generally understood that relevance is a multi-level and multi-dimensional relationship between an information need and an information object (Borlund, 2003; Ingwersen & Jarvelin, 2005; Saracevic, 2007); Vakkari, & Hakala 2000). An information object (hereafter document) may meet the dimensions (hereafter themes) of an information need, based on a work task context, to different degrees.

In information retrieval (IR), relevance has a major role in the evaluation of retrieval methods and systems (Croft et al., 2010). Until the end of the past millennium, the state-of-the-art (SOTA) experimental IR evaluation employed binary mono-dimensional relevance judgements: a document either is wholly relevant to an information need (or topic) or not at all relevant thereby neglecting both the multi-level and the multi-dimensional character of document relevance. Around 2000, Sormunen (2002) constructed the first large test collection based on graded (i.e., multi-level) relevance judgments, and Jarvelin & Kekäläinen (2000; 2002; Kekäläinen & Jarvelin, 2002) proposed appropriate metrics for IR evaluation based on graded assessments. Keskustalo *et al.* (2008) and Ferro *et al.* (2016) proposed metrics that take into account the searcher's displeasure for encountering non-relevant documents. None of the above dealt with multidimensionality of relevance. One focus of the present paper is the question "How to deal with multidimensional and graded relevance assessments in IR evaluation?"

Search algorithms produce ranked lists of documents. Successive documents in the ranking may provide similar content that does not add up the information already given to the user in the preceding documents. Also the evaluation metrics neglect document content overlap and naively assume gains piling up as the searcher examines the ranked list at greater length. Consequently, another focus of the present paper is the question "How to deal with document content overlap (beyond duplicate recognition) in IR evaluation?"

The usability of a document for a person-in-need also depends on other factors beyond relevance. A fully relevant (topically – see Ingwersen & Jarvelin, 2005) document may be useless for other reasons, e.g., due to its language, age, preferences, or the level of expertise or world view assumed on behalf of the reader. We call these factors as document usability attributes. While successive documents may add up to the content cumulated for each search task theme, save for overlaps, attributes do not have such a cross-document cumulation property. Each attribute forms its own dimension. The third focus of the present paper is therefore the question "How to deal with usability attributes, and how to combine this with multidimensional relevance assessments in IR evaluation?"

The fourth focus of the present paper is the question "How to define a formal model, which deals with multidimensional graded relevance assessments, document overlaps, and document usability attributes in a coherent single framework serving IR evaluation? A further desired feature is to be able to derive SOTA IR evaluation models as special cases of the proposed formal model."

Newell (1969) proposed that the definition of a method requires three components: the domain D of the method, the algorithm A, and the justification J. D specifies the kind of tasks that the method is valid for. The algorithm A specifies the inputs, processing, and outputs of each step of the method. Finally, the justification J contains arguments to convince the user of the method that it is rational to believe that the method works and delivers valid and reliable results. A proposal for a new method entails and requires pointing out novelty either in the tasks, algorithm and/or justifications. In the comparison of old and proposed new methods one may use criteria such as simplicity, accuracy, scope, systematic power, explanatory power, fruitfulness, and validity as discussed in Järvelin and Wilson (2003). Because a metric / test collection is aka method, we follow Newell's proposal in the definition of the proposed new metric.

It is generally known that when some activity is developed through repeated evaluations over time, measured qualities tend to become ends of evaluation and set the foci of development. We call this the principle of WYSIWYG (what you see is what you get) in design. Its corollary also works here, the principle of WYDSIWYDU (what you don't see is what you don't understand). Therefore one should set easy evaluation tasks and their naïve effectiveness metrics aside and commit to a more formidable evaluation challenge with associated metrics – even if they prove one's search algorithms are ineffective and one's approach narrow and unprepared.

We use the Cumulated Gain (CG) family of IR evaluation metrics by Jarvelin & Kekäläinen (2002) as the foundation of the new metrics which we call MDCU for multi-dimensional cumulated utility.

## 2. The task of the MDCU test collection framework

### 2.1 Goals and requirements

Experimental IR is traditionally based on test collections (Ingwersen & Järvelin, 2005; Sanderson, 2010). A test collection consists of a typically large document collection, a set of search tasks (or topics), and a relevance corpus (often called *qrels*) indicating, which documents are relevant to which topics. Relevance is mono-dimensional and often binary (Harman & Voorhees, 2006), sometimes graded (Sormunen, 2002). A standard search task representation consists of a short title T, a couple-of-sentences-long description of the information need D, and a narrative N sharpening the relevance criteria. Competing search mechanisms are tested for their ability to find, for each search task, the relevant documents and rank then in descending order of relevance.

The limitations of this widely used approach include the simplistic handling of relevance (mono-dimensional and binary), commitment to topical relevance (neglecting task-based and situational relevance), assumption on document independence regarding relevance (repeated information fully rewarded), assumption on interaction with search results (late-ranked documents are as valuable as first-ranked if equally relevant). These limitations have been criticized from the standpoint of user-oriented IR research (e.g., Froelich, 1994; Schamber, 1994; Harter & Hert, 1997; Borlund & Ingwersen,1998). In many empirical studies, real searchers have applied a wide spectrum of subjective and dynamically changing relevance criteria in assessing documents retrieved (e.g., Barry, 1994; Barry & Schamber, 1998; Greisdorf, 2003; Kekäläinen & Järvelin, 2002; Maghlaughlin & Sonnenwald, 2002; Rieh, 2002; Spink & Greisdorf, 1998). Overcoming these limitations requires that the IR evaluation framework contains explicit, and preferably manipulatable, representations of task, searcher, and situational features of information interaction behind the searching under evaluation. In addition, evaluation metrics that take these features into account, are needed.

Belkin (2016) proposed moving from relevance to usefulness as the main criterion for evaluation of IR system performance. That would mean evaluating an IR system for its capability in helping people to achieve a motivating task/goal. This would require the development of an empirically based taxonomy of such tasks/goals. Byström & Järvelin (1995), Kumpulainen & Järvelin (2010), Li & Belkin (2008), Saastamoinen

& Järvelin (2017) and Vakkari (2016) provide ideas for building such taxonomies and inferring requirements for searching that would be helpful. However, these works do not propose representations nor metrics for evaluation of retrieval. Novel approaches need to be developed if new or broader evaluation scenarios are desired (Kekäläinen & Järvelin, 2002). The proposal we make is a step toward Belkin's usefulness, covering search evaluation but not other aspects of information interaction (see Järvelin & al., 2015)

The simulated work task situations (Borlund, 2016; Ingwersen & Jarvelin, 2005) aim at bringing task-based and situational aspects into the evaluation of interactive IR (IIR). The approach has been used widely to bring realism into IIR evaluation with a main impact on experimental design and the interactive searcher's behavior. This approach affected the representation of search tasks and aimed at guiding a test subject's behavior. However, there were no new proposals regarding how to deal with the representation of multi-dimensional relevance, document overlap, and attributes in search result evaluation.

Many metrics are available for measuring the quality of the search results per search task and across them. The traditional metrics include recall (R), precision (P), precision at cutoff, mean average precision (MAP), see (Sanderson, 2010). They are suitable for binary relevance in mono-dimensional relevance contexts. Generalizations exist, like generalized recall gR, precision gP, and mean average precision (gMAP), see (Kekäläinen & Järvelin, 2002; Robertson *et al.*, 2010), and metrics "born-graded", like the Cumulated Gain (CG) based metrics (Jarvelin & Kekäläinen, 2002) but these are still for mono-dimensional relevance, and do not.handle document overlaps and usability. One contribution of the interactive track of TREC was the notion of *instance relevance,* intended to solve the problem of retrieving overlapping information in the interactive user tests (Over 2001). A limitation of this approach was that the search tasks must focus on unique "countable instances" – which tasks are not typical.

The INEX retrieval experiments had graded topical assessment and filters pointing out the XML structural elements where the required answer should be found (INEX, 2014; Stanford, 2022). The evaluation metric credited each retrieved answer element for its topical match combined with structural exactness in relation to the search task. One may claim that by applying suitable structural filters the INEX evaluation metric had some support for multi-dimensional topical relevance and attributes. Nevertheless, the scoring criteria for structure only took into account how comprehensively each search result contained the answer. There was no explicit approach for integrating multiple independent attributes into evaluation. Fuhr and Roelleke (1997) proposed the probabilistic relational model as a generalization of the standard relational model. It offers a well-defined mechanism for handling probabilistic inference on both textual content and uncertain database attribute values. This can be seen as a scoring mechanism, in particular for the multidimensional attributes, while it does not directly support handling of multiple themes and document overlaps.

Document overlap determination for cumulated utility calculation is a tricky problem. Traditional mono-dimensional and one document at a time relevance scoring ignores overlaps. It does not encourage ranking a minimal number of documents that collectively provide an as comprehensive response to the search task as possible to the top ranks of the search result. Instead, it rewards retrieval models for clotting the result with redundant answers. Maximising evaluation metric values replace original more fundamental goals. There are studies analysing the overlap between search results from different sources or produced by different types of queries or document representations (Katzer *et al.,* 2017; Saracevic *et al.*, 1988). The focus there is on retrieving the *same* document by different means – whereas we are interested in *different* documents contributing complementary content on the same themes compared with other (sets of) document(s). Studies of clustering and text classification (Mirończuk & Protasiewicz, 2018) focus on textual similarity between documents while the present challenge is finding thematically similar documents that are textually as different as possible until the thematic utility is maximised for all themes. A further problem is that any sequence of retrieved documents should be possible to evaluate. Modern test collection sizes do not allow intellectual assessment in this case.

The document usability aspects of documents have not received as much attention in IR literature as relevance. We do not know of any systematic approach to handle usability attributes – their selection, measurement, or calculation of evidence. Typically, selection is based on ad hoc criteria on document metadata fields, measurement is as given in the document metadata, and the evidence combined by Boolean functions. There is little discussion on capitalizing on other data available in the search environment. A noteworthy proposal is the "information nutrition label" (Fuhr & al., 2017) that can be used to describe document usability. Among others, such a label can provide information on language, genre, factuality, virality, opinion, controversy, authority, technicality, and topicality of a document. Some attributes may be general, automatically computable, and hold for all search tasks, some depend on a particular task / searcher combination.

Table 1 compares traditional IR evaluation with the multidimensional approach proposed in this paper.

The task of the proposed framework is to score searchers' search results in IR tasks when document relevance is multidimensional, the contributions of individual documents in each theme may overlap, and document usability attributes affect scoring. Therefore the framework must:

- offer a document representation for assessing (a) the contribution of each document to the multi-dimensional topical information base in the test collection and (b) usability attributes. Within a document, the scoring of themes and attributes is independent.
- Be computable, allowing overall scoring of document utility across their thematic relevance and usability attributes. The scoring must take into account the position of each document in a search result and downgrade its score if the document is ranked low and if the preceding documents have contributed plenty of information on a particular theme. The overlap assessment between documents must avoid intellectual assessment of all possible document sequences that may appear in search results; the overlaps may be partial within and across content themes.
- reward for good attribute values (document's usability in search task context); downgrade for inappropriate attribute values.

- support systematic experimentation with characteristics of tasks, searchers, and situations.

**Table 1.** Comparison of traditional and the proposed MDCU evaluation frameworks.

| **Task feature** | **Approach to IR evaluation task** | |
|---|---|---|
| | **Traditional** | **MDCU** |
| **Relevance degrees** | Binary | Graded |
| **Relevance dimensions** | One (topical) | Multiple task-based themes, multiple attributes |
| **Relevance judgment, cost** | Human, €€ | Human, €€€ |
| **Relevance basis** | “There is at least one relevant sentence in the document text” | “The amount of relevant text correlates with relevance degree per theme Usability attributes may be observed/compututed.” |
| **Relevance features: content / structure / attributes** | Yes / no / no | Yes / no /yes |
| **Result score range** | [0, ∞), [0, 1], % | [0, ∞), [0, 1], % |
| **Overlap handling** | Not supported | Multidimensional scoring offers some possibilities |
| **Topic design cost** | € | €€€ |

## 2.2 Basic document representation

We propose a three-component representation of documents for the MDCU framework: content (including metadata), theme relevance list, and usability attribute list. A list of documents, e.g., a search result, relevance corpus, or relevance feedback list, may thus be presented in table form, see Table 2. It presents an abstract six document list, d1 – d6, with the content component suppressed. Document relevance is represented for the imaginary search task through 4 themes, Theme1 – Theme4, with integer values ranging from 0 (non-relevant) to 3 (highly relevant), and three imaginary attributes, Attr1 – Attr3, with real values for usability between 0 to 1. The content themes are determined by the (search) task, and the relevance scores by the themes and the documents. Here we follow Sormunen’s (2002) model of four relevance levels due to its popularity, but any other scaling could be used. The usability attributes are determined by the searcher and the search task, and the usability scores by the attributes and the documents. In principle even a highly relevant document may have zero usability.

Let us assume that we have somehow constructed a representation for the task and the searcher characteristics through such themes and attributes, and further, that we have obtained the descriptive numbers – we will discuss that later. We may now focus on how the MDCU framework works once we have the numbers. We will first consider the formal (i.e., content-free) characteristics of the framework: a) how various aspects can be represented, b) how variables interact in combining evidence for scoring, and c) how the framework behaves in abstracted situations.

**Table 2**. A simple model for representing multidimensional relevance quantity, with range {0, 1, 2, 3}, together with attribute usability scores in the range [0,1].

| | **Degree of theme relevance** | | | | | **Usability attribute scoring** | | | |
|---|---|---|---|---|---|---|---|---|---|
| **Document id#** | Theme1 | Theme2 | Theme3 | Theme4 | … | Attr1 | Attr2 | Attr3 | … |
| **d1** | 0 | 1 | 3 | 2 | … | 1,0 | 1,0 | 1,0 | … |
| **d2** | 2 | 0 | 0 | 2 | … | 0,9 | 0,7 | 0,9 | … |
| **d3** | 1 | 0 | 2 | 0 | … | 1,0 | 0,9 | 1,0 | … |
| **d4** | 0 | 0 | 3 | 1 | … | 0,8 | 0,9 | 0,7 | … |
| **d5** | 1 | 2 | 0 | 2 | … | 1,0 | 1,0 | 1,0 | … |
| **d6** | 0 | 0 | 0 | 2 | … | 1,0 | 0,8 | 1,0 | … |
| **…** | … | … | … | … | … | … | … | … | … |

## 2.3 The components of the framework

A test collection has several components, described next with special features required by the MDCU framework highlighted. These are described in the following insert.

| **Component** | **Description** |
|---|---|
| Document collection DC | The documents have a textual (or media) component and 0 – *n* metadata fields, which may have explicitly stored or functionally produced values, e.g., *lang*(doc) = "French". |
| Search tasks ST | The search tasks are represented in the traditional way as a TDN (Tile – Description – Narrative) structure. Type one search tasks TDN-1 are all-verbal, including requirements on attributes, e.g., "scholarly articles in French", while TDN-2 tasks have a formal way of expressing such conditions, e.g., *type*(doc) = "scholarly" & *lang*(doc) = "French". The TDN-1 tasks bring more uncertainty to the query construction process and result scoring.<br><br>The N-component contains the description of each theme and attribute of the search task, explaining the relevance criteria and requirements for each degree of relevance. It is beneficial, if semantically indicative vocabulary is collected for each theme – this could be organized as an ontology (Jarvelin *et al.*, 2001). |
| Relevance corpus RC | RC gives for all documents assessed for a search task a representation shown in Table 2. The rest of DC is implicitly not relevant / not usable.<br><br>Themes are assessed in documents independently numerically; theme scoring may have different ranges; here using with no loss of generality integer values {0, 1, 2, 3} from not relevant to highly relevant; note that in principle the assessment scales may vary between themes. The contents of a given theme may overlap across documents.<br><br>Document attributes are assessed in independently and numerically. Attribute scoring may have different ranges; here assumed with no loss of generality real |
| | values [0, 1]. The set of attributes of a given document interact whereas the values of a given attribute do not interact across documents. For example, if the ST contains the requirement *type*(doc) = "scholarly" & *lang*(doc) = "French", finding one, say $d_4$, meeting this requirement does not affect scoring the attributes of the following documents $d_5$ etc. |
| Scoring the search result | Theme *overlaps* between documents are treated formally: if two documents are relevant to the same theme, their expected theme overlap is estimated formally by a parametric function, avoiding intellectual analysis of overlap (of every possible combination of documents). If two documents are relevant to the same theme with scores s1 and s2, their joint score is between *max*(s1, s2) and s1+s2, depending on overlap. The contribution of s2 over s1 is discounted by a logarithmic factor.<br><br>*Cumulated gain* by themes: Correspondingly, if $CTR_j$ is the theme relevance mass cumulated in the result ranking up to rank *j* and *sj1* the theme relevance of the next document, their joint score *js* is in the range $max(CTR_j, sj1) \leq js \leq CTR_j + sj1$. Even with high CTR a new theme-relevant document contributes some to the overall theme gain score.<br><br>A document's *total utility score* is the sum of its discounted theme relevance scores downscaled by its usability attribute factors. The usability attributes downgrade document theme relevance scores uniformly through document's gross attribute usability, calculated as the product of its attribute usability factors. The relevance scores are multiplied by this.<br><br>The *cumulated utility* of a ranked search result at rank *j* is the sum of document $d_1, \ldots, d_j$ utility scores. However, the contribution of each document $d_2, \ldots, d_j$ may be less than its utility score taken independently, when overlaps are discounted.<br><br>The *ideal ranking* is based on the relevance corpus where first comes the document with the greatest total utility and next (progressively) the one that contributes most – overlaps discounted – to the already cumulated relevance.<br><br>The *normalized cumulated utility* is obtained by dividing the actual one by the ideal one for each rank. Its value at any rank is in the range [0, 1]. |

# 4. Defining it formally

## 4.1. Document representation and overlap

We shall now define some notational conventions and representations formally.

*Notational convention 1:* Let $e_1, e_2, \ldots, e_n$ be some elements, i.e., integers, character strings, or constructions thereof. The sequence $E = <e_1, e_2, \ldots, e_n>$ is *a tuple* of these elements. E[*i*] denotes its *i*th element; the operator $<head \mid\mid tail>$ splits a sequence into the *head* element and the *tail* sequence. For example, $E = <e_1 \mid\mid E'>$ gives *head* = $e_1$, and $E' = <e_2, e_3, \ldots, e_n>$

*Notational convention 2:* Let $e_1, e_2, \ldots, e_n$ be any elements, i.e., integers, character strings, or constructions, tuples are concatenated by $t = <e_1, e_2>$ and more generally $\times_{i=1,\ldots,n}\, e_i = <\ldots<e_1, e_2,>\ldots,e_n> = <e_1, e_2,\ldots,e_n>$. A null tuple of length *n* is constructed by *null-struct*(*n*) = $<e_1, e_2, \ldots ,e_n>$, where all $e_i = 0$. We extend the set membership operator '$\in$' to tuple membership '$\in^*$'. Given a tuple $t = \langle e_1, e_2, \ldots, e_n \rangle$, the expression $e \in^* t$ is true, if $\exists\, i : t[i] = e$. The length of $t$ is $length(i) = n$.

*Notational convention 3:* Let $e_1, e_2, \ldots, e_n$ be any elements considered atomic, structures are constructed by $s = (e_1, e_2)$ and more generally by $(\ldots(e_1, e_2), \ldots, e_n) = (e_1, e_2, \ldots, e_n)$.

*Notational convention 4:* Let $E = \langle e_1, e_2, \ldots, e_n \rangle$ and $F = \langle f_1, f_2, \ldots, f_n \rangle$ be any numeric value tuples. Their pairwise sum is $E \oplus F = \langle e_1+f_1, e_2+f_2, \ldots, e_n+f_n \rangle$. Likewise for product, denoted $E \otimes F$, and division, denoted $E \varnothing F$.

*Notational convention 5:* Let $dc_1, dc_2, \ldots, dc_n$ be any documents with content and structure here left unanalyzed (i.e., considered atomic). The document collection is the set $DC = \{ dc_1, dc_2, \ldots, dc_n \}$.

*Definition 1:* Document representation is a triple $d = (dc, TRel, Attr) = (dc, \langle r_1, r_2, \ldots, r_n \rangle, \langle a_1, a_2, \ldots, a_m \rangle)$, where dc is the content of the document (here left unanalyzed), $TRel = \langle r_1, r_2, \ldots, r_n \rangle$ is a tuple of theme relevance scores for a search task consisting of themes 1 … *n* in dc, and $Attr = \langle a_1, a_2, \ldots, a_m \rangle$ is a tuple of attribute usability factors for a search task consisting of usability factors 1 … *m* in dc.

*Definition 2:* Let d = (dc, TRel, Attr) be a document representation. We refer to the components of d by *cont*(d) = dc, *trels(*d) = TRel, and *attrs(*d) = Attr. We also allow the shorthand notations d[1] = *cont*(d), d[2] = *trels(*d), and d[3] = *attrs(*d).

*Definition 3:* The relevance corpus RC for a given search task ST is a set of document representations $RC_{ST} = \{d_1, d_2, \ldots, d_k\}$ for *k* documents $d_{1\ldots k}$ that have been explicitly assessed for ST. For all $d \in RC_{ST}$ the condition *cont*(d) $\in$ DC holds, i.e., the documents represented in RC belong to the document collection DC. For any unassessed documents dc $\in$ DC but $\neg\exists$ d $\in$ RC : *cont*(d) = dc, the document representation is the tuple d = (dc, *null-struct*(*n*), *null-struct*(*m*)) when ST has *n* themes and *m* usability factors.

*Definition 4:* Given a query Q and document collection DC, a search result is a tuple of document representations $SR = \langle d_1, d_2, \ldots, d_k \rangle$. We leave the definition of the retrieval function *retrieve*(Q, DC) open. The condition holds: $\cup_{d \in SR} \{cont(d)\} \subseteq DC$.

*Definition 5:* The cumulated relevance representation for *n* themes is $crr = \langle r_1, r_2, \ldots, r_n \rangle$ where $r_i \in \Re$. When the quality of the search result is assessed, *crr* accumulates the gain rank by rank for each theme of the search task. Initially, before assessing the first document, *crr* = *null-struct*(*n*).

*Definition 6:* Given a document representation $d = (dc, \langle r_1, r_2, \ldots, r_n \rangle, \langle a_1, a_2, \ldots, a_m \rangle)$, all $a_i \in [0, 1]$, the document usability factor is the product of individual usability attribute factors defined as $attr\text{-}fact(d) = \prod_{a \in^* attrs(d)} a$ .

*Definition 7:* Given a cumulated relevance representation $crr = \langle r_1, r_2, \ldots, r_n \rangle$, *b* a small integer to serve as a logarithm base for controlling overlap, and a document representation d = (dc, drel, dattr) = $(dc, \langle r_1, r_2, \ldots, r_n \rangle, \langle a_1, a_2, \ldots, a_m \rangle)$, their combined relevance is:

$$contrib(crr, \text{drel}, \text{dattr}) = \times_{i=1,\ldots,n} (crr[i] + v * \text{ drel}[i] / max(1, log_b\, crr[i]))$$

This discounts the contribution of d on themes *i*, *i* = 1 … *n*, by the logarithm of the cumulated relevance mass for the theme *i* (minimum of 1) and by the document usability factor *v* = *attr-fact*(d). The result is a new cumulated relevance representation *crr*' with components $r'_i$ incremented by the contribution of corresponding d components. Note that discounting begins only when the cumulated relevance mass exceeds the base of the logarithm *b*. By setting *b* small ( ≤ 1.5) a high overlap is modeled. As an example, consider the following document representations $d_1$ and $d_2$ with four themes and three attributes, and $crr = \langle 0, 0, 0, 0 \rangle$. The themes have individual relevance scores 0 … 3, the overlap parameter *b* = 2, and the attribute usability factors in the range [0 ,1].

$d_1 = (\text{dcont1}, \langle 0, 1, 3, 2 \rangle, \langle 1, 1, 1 \rangle)$ whence $v_1 = attr\text{-}fact(d_1) = 1$

$d_2 = (\text{dcont2}, \langle 2, 0, 0, 2 \rangle, \langle 0.9, 0.7, 0.9 \rangle)$ whence $v_2 = attr\text{-}fact(d_2) = 0.57$

Now $contrib(crr, d_1[2], d_1[3]) = crr' = contrib(\langle 0, 0, 0, 0 \rangle, \langle 0, 1, 3, 2 \rangle, \langle 1, 1, 1 \rangle) = \langle 0, 1, 3, 2 \rangle$,

which is the relevance scoring of $d_1$ – a document does not overlap with itself and $v_1 = 1$

but $contrib(crr', d_2[2], d_2[3]) = crr'' = \langle 1.14, 1.0, 3.0, 3.13 \rangle$,

where *crr''*[1] = 1.14 due to $v_2$, < 1, while *crr''*[2] and *crr''*[3] retain their scores earned already in *crr'*. Finally, *crr''*[4] = 3.13 estimates the overlap of theme four enough for reducing $d_1[2]\,[4] + d_2[2][4]$ by almost one point from 4 to 3.13.

*Definition 8:* Given a search result SR its rank-wise cumulated relevance is

*cum-rel*(*crr*, SR, OSR) =

OSR if SR = < >

*cum-rel*( crr ⊕ ddr, SR', <OSR, < d[1], ddr, d[3]>>),

otherwise

where SR = <d || SR'> and ddr = *contrib*(*crr*, d[2], d[3]),

In the first call of *cum-rel* the operand *crr* = *null-struct*(*n*), where *n* = *length*(d[2][1]), and OSR = <>.

Table 3 shows the major calculations based on the data in Table 2 and taking them as a ranked search result in top-down order. Column 2 shows the independent overall document score (from Table 2). The columns under Themes report the relevance per theme for each document taking into account the estimated overlap (*b* = 2) when documents are accessed in the given order. Columns 7 and 8 give the contribution across themes per document and the usability factors. The remaining two columns show the document score with the usability factors applied and the cumulative relevance score for the SERP. Note that *cum-rel*(*null-struct*(4), SERP, <>) = *serp-trel* = <3.63, 3.00, 5.696, 6.242> and $\sum_i$ *serp-trel*[*i*] = 18.57 [N.B. this score does not include usability discount]. We can read from this scoring, that the SERP is strongest vs. weakest on themes four vs. two. If the overlap parameter is set at $b$ =1.1 for modelling high overlap of documents, we get *serp-trel* = <2.26, 3.00, 3.42, 2.81> and $\sum_i$ *serp-trel*[*i*] = 11.49.

**Table 3**. A simplified SERP with scoring components for multidimensional relevance – based on Table 2 data. Relevance scoring {0, 1, 2, 3}, usability factors [0, 1], and $b$ = 2.

| **SERP** | **Orig** | **Themes** | | | | **Contrib** | **UsabF** | **Doc** | **Cum** |
|---|---|---|---|---|---|---|---|---|---|
| **Doc id#** | **TotRel** | $r_1$ | $r_2$ | $r_3$ | $r_4$ | $\sum r_i$ | $v_i$ | **Score** | **Rel** |
| $D_1$ | 6 | 0,000 | 1,000 | 3,000 | 2,000 | 6,000 | 1,000 | 6,000 | 6,00 |
| $D_2$ | 4 | 2,000 | 0,000 | 0,000 | 2,000 | 4,000 | 0,567 | 2,268 | 8,27 |
| $D_3$ | 3 | 1,000 | 0,000 | 1,262 | 0,000 | 2,262 | 0,900 | 2,036 | 10,30 |
| $D_4$ | 4 | 0,000 | 0,000 | 1,434 | 0,500 | 1,934 | 0,504 | 0,975 | 11,28 |
| $D_5$ | 5 | 0,631 | 2,000 | 0,000 | 0,922 | 3,553 | 1,000 | 3,553 | 14,83 |
| $D_6$ | 2 | 0,000 | 0,000 | 0,000 | 0,820 | 0,820 | 0,800 | 0,656 | 15,49 |
| **Sum** | 24 | 3,631 | 3,000 | 5,696 | 6,242 | 18,569 | … | … | … |

While the theme, utility and overlap based document and SERP scoring gives important data on query performance, a problem lies in comparability across search tasks because, as in the case of Cumulated Gain (Järvelin & Kekäläinen, 2002), the scoring is not normalized. To avoid the problem, we define the ideal and normalized multi-dimensional scoring with overlaps.

## 4.2. Ideal ranking – dynamic overlaps

The ideal ranking for a search task ST and relevance corpus RC is the best possible ordering of documents in accruing gain under the requirement of avoiding rewarding for overlapping results. In principle, one always chooses for the next rank the document of highest contribution among the remaining ones in RC. The face values of themes must not be used as such because of the already cumulated theme content. At each step in the construction of the ideal ranking we choose the document offering the greatest residual contribution.

*Definition 9:* Given a relevance corpus RC = { $d_1$, $d_2$, …, $d_k$}, its ideal ranking is

*ideal-ranking*(RC, *crr*) =

<id || *ideal-ranking*(RC – {d}, *cum-rel*(*crr*, <d>))>,

when d ∈ RC ∧ ∀ d’ ∈ RC: *contrib*(*crr*, d) ≥ *contrib*(*crr*, d’), if |RC| ≥ 2

<id>, otherwise

where id = <d[1], *cum-rel*(*crr*, d), d[3]>

In the first call of *ideal-ranking* the operand *crr* = *null-struct*(*n*), where *n* = *length*(d[2]). The construction of the ideal ranking is computationally heavy if the relevance corpus is large. It must also be reconstructed when the document set, their themes or attributes are modified.

**Table 4 (a)**. The relevance corpus RC', 10 documents. Relevance scoring {0, …, 3}, usability factors [0, 1].

| RC' | Degree of theme relevance | | | | | Usability attribute scoring | | | |
|---|---|---|---|---|---|---|---|---|---|
| **Document id#** | **The-1** | **The-2** | **The-3** | **The-4** | **…** | **Attr1** | **Attr2** | **Attr3** | … |
| **d1** | 0,00 | 1,00 | 3,00 | 2,00 | … | 1,00 | 1,00 | 1,00 | … |
| **d2** | 2,00 | 0,00 | 0,00 | 2,00 | … | 0,90 | 0,70 | 0,90 | … |
| **d3** | 1,00 | 0,00 | 2,00 | 0,00 | … | 1,00 | 0,90 | 1,00 | … |
| **d4** | 0,00 | 0,00 | 3,00 | 1,00 | … | 0,80 | 0,90 | 0,70 | … |
| **d5** | 1,00 | 2,00 | 0,00 | 2,00 | … | 1,00 | 1,00 | 1,00 | … |
| **d6** | 0,00 | 0,00 | 0,00 | 2,00 | … | 1,00 | 0,80 | 1,00 | … |
| **d7** | 0,00 | 0,00 | 0,00 | 0,00 | … | 0,00 | 0,00 | 0,00 | … |
| **d8** | 1,00 | 1,00 | 1,00 | 0,00 | … | 0,30 | 1,00 | 1,00 | … |
| **d9** | 0,00 | 0,00 | 0,00 | 2,00 | … | 0,90 | 0,90 | 0,90 | … |
| **d10** | 3,00 | 3,00 | 3,00 | 1,00 | … | 1,00 | 1,00 | 1,00 | … |

**Table 4 (b)**. The ideal ranking based on RC' Relevance scoring {0, 1, 2, 3}, utility factors [0, 1], and *b* = 1.5.

| Ideal iOAr | Degree of theme relevance | | | | Face | Disc | Attrib | Doc | |
|---|---|---|---|---|---|---|---|---|---|
| **Document id#** | **The-1** | **The-2** | **The-3** | **The-4** | **Value** | **Contrib** | **Factor** | **Score** | **iOAr** |
| **rank1 d10** | 3,00 | 3,00 | 3,00 | 1,00 | 10,00 | 10,00 | 1,00 | 10,00 | 10,00 |
| **rank2 d1** | 0,000 | 0,369 | 1,107 | 2,000 | 6,00 | 3,48 | 1,00 | 3,48 | 13,48 |
| **rank3 d5** | 0,369 | 0,668 | 0,000 | 0,738 | 5,00 | 1,77 | 1,00 | 1,77 | 15,25 |
| **rank4 d3** | 0,334 | 0,000 | 0,574 | 0,000 | 3,00 | 0,91 | 0,90 | 0,82 | 16,07 |
| **rank5 d2** | 0,619 | 0,000 | 0,000 | 0,615 | 4,00 | 1,23 | 0,57 | 0,70 | 16,77 |
| **rank6 d4** | 0,000 | 0,000 | 0,861 | 0,369 | 4,00 | 1,23 | 0,50 | 0,62 | 17,39 |
| **rank7 d6** | 0,000 | 0,000 | 0,000 | 0,522 | 2,00 | 0,52 | 0,80 | 0,42 | 17,81 |
| **rank8 d9** | 0,000 | 0,000 | 0,000 | 0,489 | 2,00 | 0,49 | 0,73 | 0,36 | 18,16 |
| **rank9 d8** | 0,277 | 0,291 | 0,237 | 0,000 | 3,00 | 0,80 | 0,30 | 0,24 | 18,40 |
| **rank10 d7** | 0,000 | 0,000 | 0,000 | 0,000 | 0,00 | 0,00 | 0,00 | 0,00 | 18,40 |

Table 4 (a) gives a relevance corpus RC' for 10 documents with d1 – d6 being the same as in Table 2. Table 4 (b) presents their ideal ranking. The contributions of each document on each theme and in total are given, both the independent value and discounted by the estimated overlap based on a moderate parameter *b* = 1.5. The three columns on the right contain the attribute factor *v*, the document score at each rank and the cumulative gain at each rank.

Note that it is a simplifying design choice to apply a single overlap parameter *b* and a single attribute factor *v* across all themes. Experiments may indicate that multiple coefficients are worth the additional complication.

### 4.3 Normalized gain

The possible cumulated gain of search task SERPs varies greatly depending on the size of the relevance corpus, assumed document overlaps, and the stringency of the attribute factors. The gains across search tasks are not comparable and some search tasks may dominate the overall picture (Jarvelin & Kekäläinen, 2002). When comparability across search tasks is needed, one may normalize the search result gains relative to the ideal result. The normalized gain may be calculated in a simple way, given a search result SR and a relevance corpus RC with its ideal ranking is IR = *ideal-ranking*(RC, *null-struct*(*n*)). Both are tuples of document representations d = (dc, TRel, Attr) where each theme relevance component is discounted for overlaps and cumulated across the ranking. It remains to extract the cumulated relevance vector for both rankings and divide the observed performance by the ideal performance rank-by-rank. This can be done for each theme separately, or any of their combinations. It can also be done for any subset of attributes. We define below functions for aggregating the theme-based relevance figures into an overall score, discounted by the attribute factors. First we define the functions *doc-gain* and *gain-vector* for extracting the cross-theme gain of a single document and constructing the gain vector based on a ranking of documents, respectively.

*Definition 10:* Let d = (dc, Trel, Attrs) be any document and R = $<d_1, d_2, \ldots, d_k>$ any ranking of *k* documents. The gain d provides is

$$doc\text{-}gain(d) = \sum_{i=1,,\ldots,n} Trel[i] * attr\text{-}fact(d)$$

The cumulated gain vector for R, based on cross-theme relevance with usability discount, is

$$gain\text{-}vector(R, k) = \;< cg_1, cg_2, \ldots, cg_k > \quad \text{where } cg_i = \sum_{j=1,\ldots,i} doc\text{-}gain(d_j)$$

For example in Table 4 (a), document d4 provides the gain *doc-gain*(d4) = (0.00+0.00+3.00+1.00) * (0.80*0.90*0.70) = 4*0.5 = 2. When this document is ranked at rank 6 in Table 4 (b), now with overlap discounts, we get *doc-gain*(d4) = (0.00+0.00+0.861+0.369) * 0.5 = 0.62. The right-most column in Table 4 (b) gives the gain vector of the sample ideal ranking, ending with the score 18.40 points at rank 10.

Finally, we define gain vector normalization as division operation between two vectors, component by component. The search result provides the numerator components $< sg_1, sg_2, \ldots, sg_k >$ and the ideal ranking the denominator components $< ig_1, ig_2, \ldots, ig_k >$.

*Definition 11:* Let IR be an ideal ranking and SR the observed ranked search result to be normalized, both of length $k$, and ig = *gain-vector*(IR, $k$) and sg = *gain-vector*(SR, $k$) the corresponding gain vectors. The normalized gain vector is:

$$\textit{norm-vector}(sg, ig) = sg \oslash ig = < sg_1/ig_1, sg_2/ig_2, \ldots, sg_k/ig_k,> .$$

Table 5 gives an example on the calculation of normalized gain vector.

**Table 5.** The calculation of normalized gain vector, given search gain vector and ideal gain vector, both based on theme relevance scoring {0, 1, 2, 3}, usability factors [0, 1], and $b$ = 1.5.

| SR | Search CG | Ideal R | iCG | nCG |
|---|---|---|---|---|
| **d1** | 6,00 | **ir1 d10** | 10,00 | 0,60 |
| **d2** | 7,80 | **ir2 d1** | 13,48 | 0,58 |
| **d3** | 8,99 | **ir3 d5** | 15,25 | 0,59 |
| **d4** | 9,63 | **ir4 d3** | 16,07 | 0,60 |
| **d5** | 12,70 | **ir5 d2** | 16,77 | 0,76 |
| **d6** | 13,16 | **ir6 d4** | 17,39 | 0,76 |
| **d7** | 13,16 | **ir7 d6** | 17,81 | 0,74 |
| **d8** | 13,46 | **ir8 d9** | 18,16 | 0,74 |
| **d9** | 13,84 | **ir9 d8** | 18,40 | 0,75 |
| **d10** | 16,84 | **ir10 d7** | 18,40 | 0,92 |

## 5. Discussion

The MDCU framework can be justified through several criteria (e.g., Järvelin & Wilson, 2003). We shall focus on its goal and scope, its behavior under various conditions, issues in building the test collection, validity and cost, support to experimentation, and integration with other areas of IR.

### 5.1. Goal and scope of the MDCU or IR evaluation

*The Goal of IR and IR Evaluation.* Traditionally, the experimental goal of IR is finding topically relevant documents for a search request. IR evaluation provides tools and metrics for assessing how well search engines succeed in this. More and more accurate are merits of better engines, the main experimental variable. We are proposing more demanding goals: IR should find as concise as possible answers to help a person overcome a problematic situation. The requests therefore may need to refer to multiple aspects of the person's task (or other engagement), the existing vs. lacking knowledge of the person, and her/his personal capabilities, competencies, and interests. Excluding these aspects means providing just lip-service to task-based evaluation. The revised goals are also relevant when IR serves as a retrieval component of a modern question answering or interactive AI system.

*The Scope of the Evaluation Framework.* We have presented an abstract framework for incorporating task, situation, and person related aspects into IR evaluation. It is abstract, because we have not specified the themes of relevance, nor the attributes, for any specific evaluation case – and do not have a test collection. However, these can be specified, for example, beginning with the sketch:

- Person: high-school student in Zurich, moderately leaning toward the green movement
- Task: essay on anti-climate change measures for classroom presentation
- Situation: initial collection of material, little prior knowledge on the subject
- Document collection: multi-lingual media archive in 5 languages in Switzerland
- Themes: polluters, pollutants, consequences, green tech, negotiations, agreements, …
- Attributes: is-feature-article, is-green-biased, is-French, is-shared-by-FB-friends, is-layman-oriented, ….

These themes and attributes contain topical, task-based, situational, and personal aspects of the case. We have shown that the framework can integrate them in a computable form for IR evaluation. Therefore the MDCU framework has broader scope than the traditional evaluation model, which does not push one to look equally broadly for reasons to success / failure. Conceptually poorer models provide a weaker basis for, and lead to more constrained analyses of, success.

*The principal difference*. The traditional IR evaluation approach focuses on the relevance of the sequence of items in the search result and provides several metrics for measuring the quality of the result. Consider Table 3 as the SERP S1 returned by a search engine though traditional goggles: Appendix 1 shows S1 collapsed into mono-dimensional average relevance and relevance score vectors in several ways -- binarization with two thresholds for P@R (precision at rank), and several cumulated gain-based vectors: iCG (ideal cumulated gain), S1 CG, iDCG (ideal discounted CG), and S1 DCG. These are illustrated in Figures 1 (a)-(b). Each metric forms a series of columns with one reading at each rank (document). While this reading tells about the gain of S1, we do not learn about the utility of S1, nor the degree of overlap, nor underrepresented themes.

Figures 1 (c)-(d) show MDCU-based scoring of SERP1. On the left, the naïve scoring is based on the aggregated relevance for each item multiplied by its usability. We can see that d4 and d8 do not score well due to low usability. On the right, the relevance dimensions are opened. We can see that the scores of individual items are based on different dimensions (e.g., d3 vs. d5), suggesting less overlap, while d6 and d9 likely are redundant and poor. Items d1 and d10 could form a better concise two-document pair than the entire SERP1. MDCU offers metrics for scoring and comparing SERP quality.

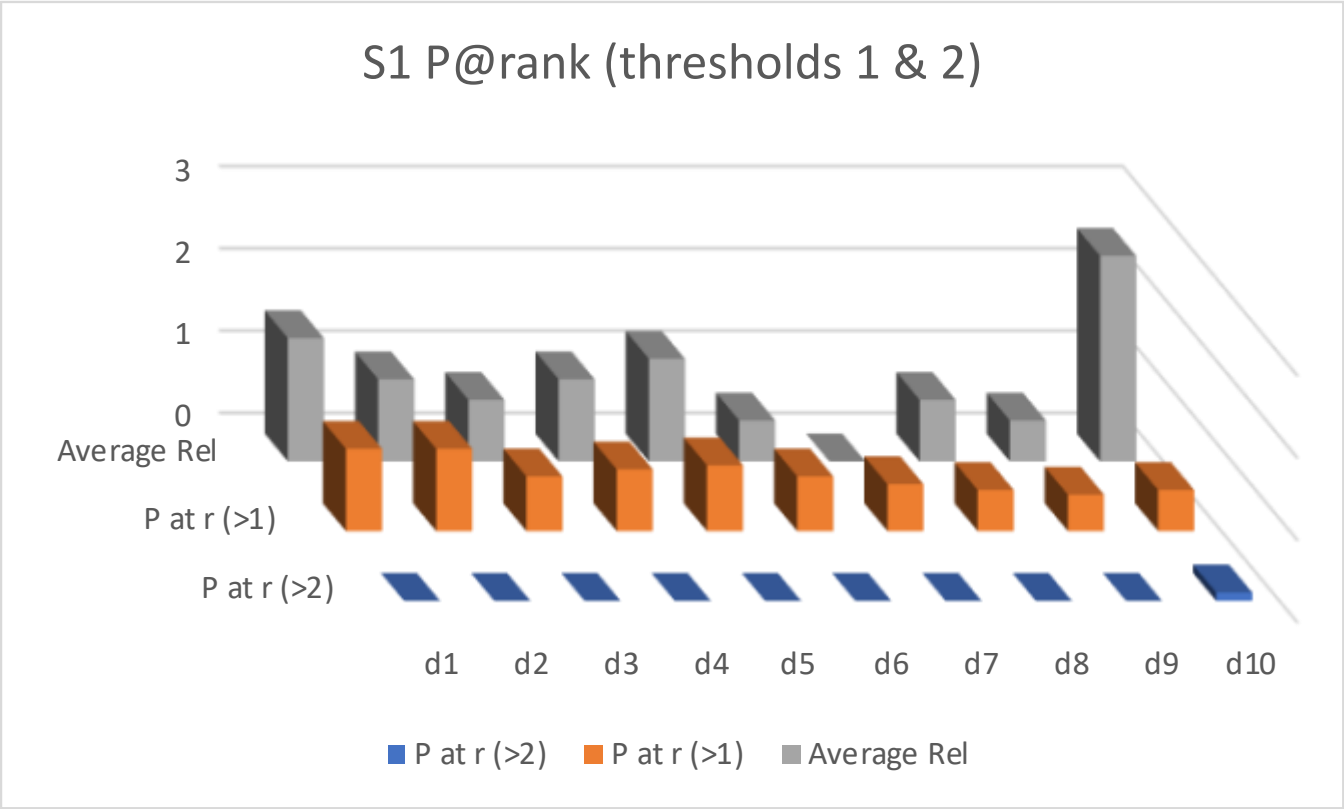

(a)

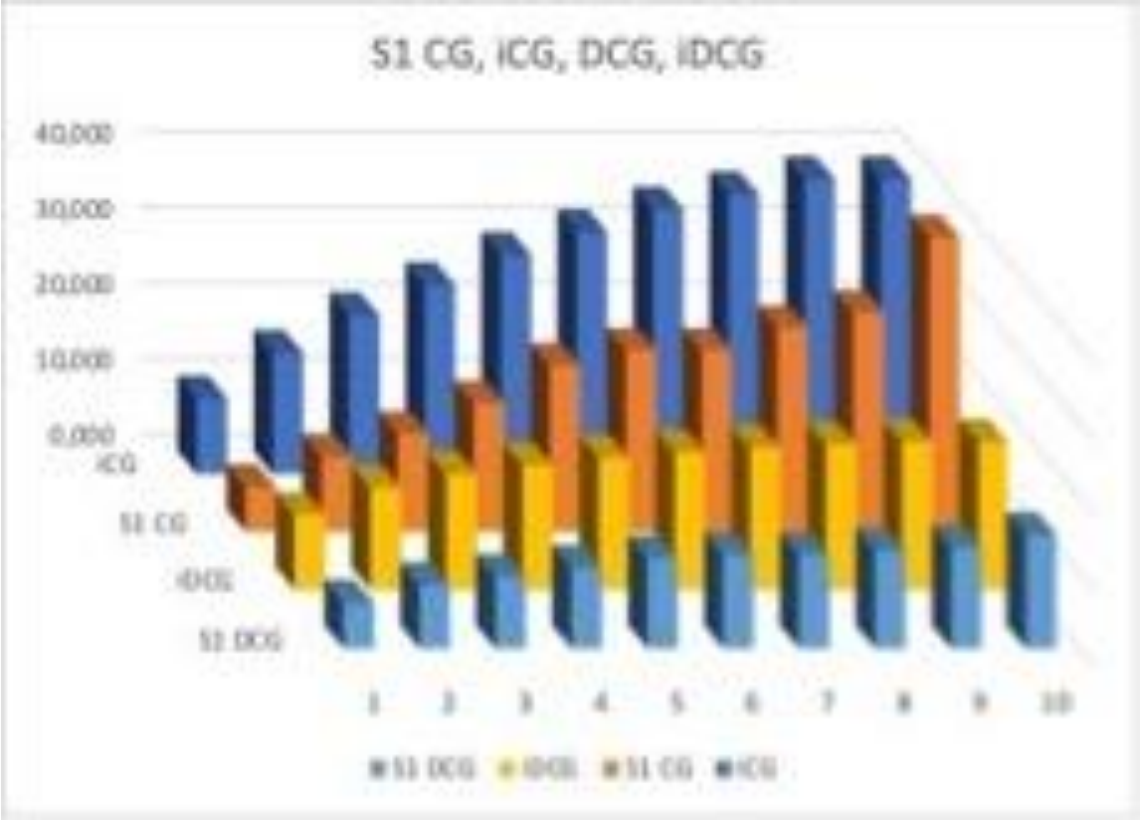

(b)

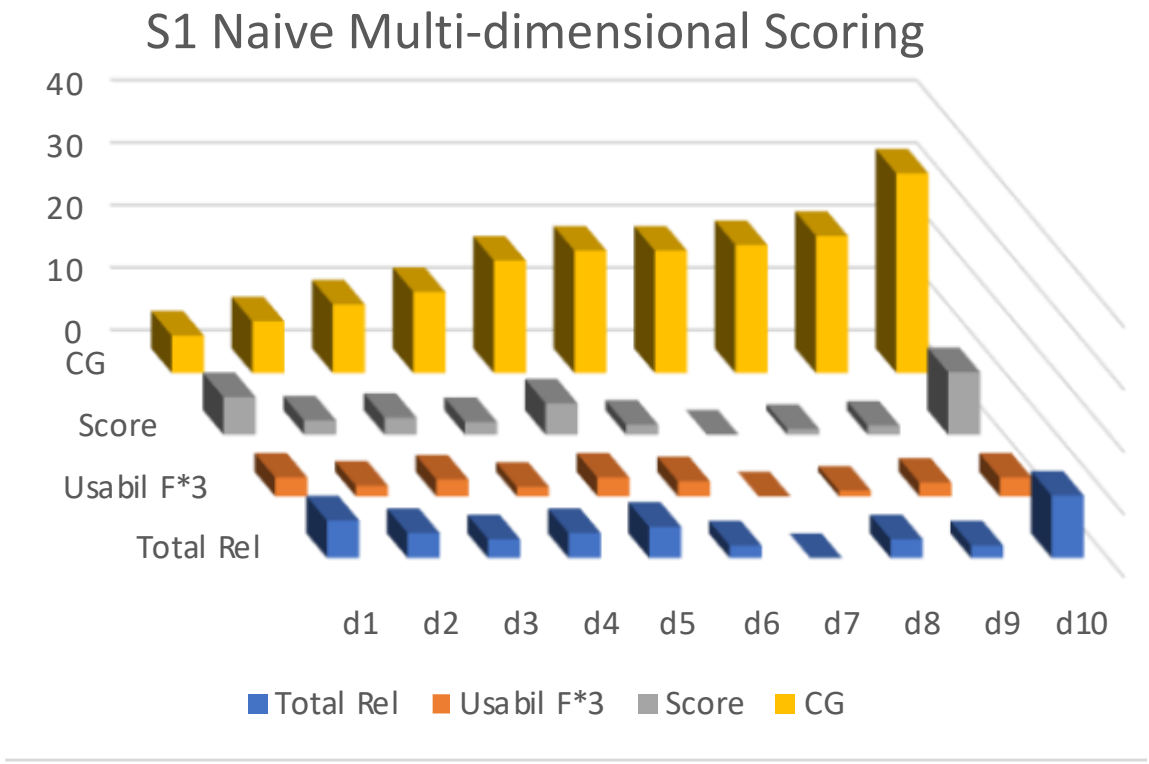

(c)

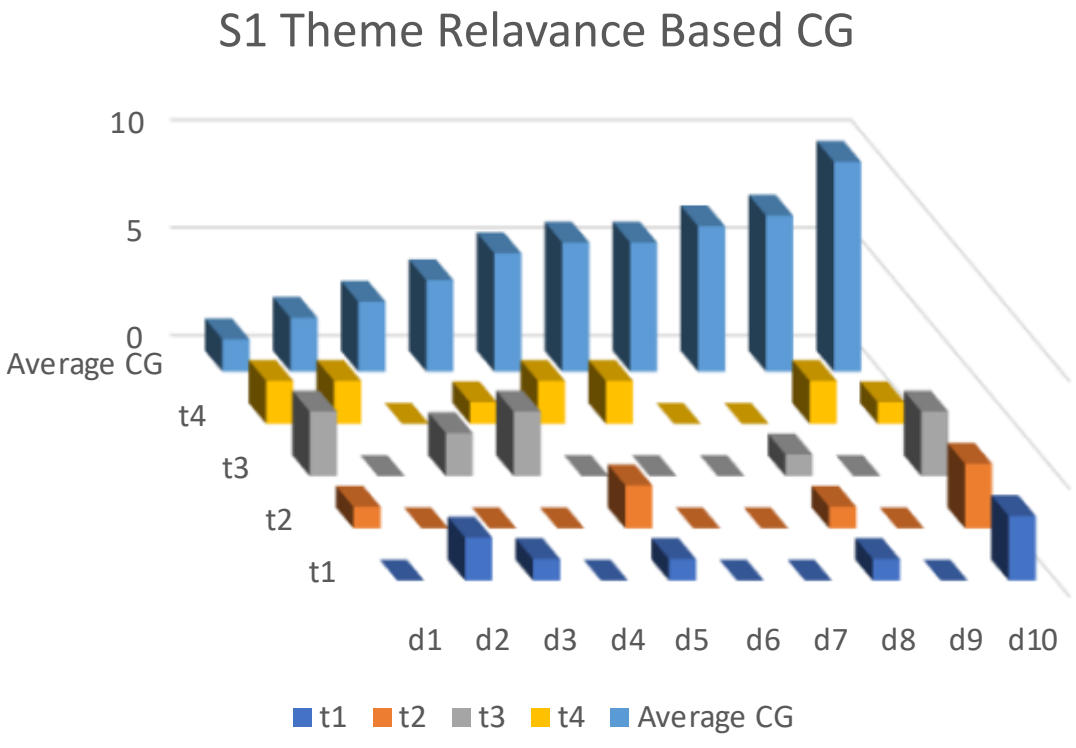

(d)

**Figure 1. (a)-(b)** Illustrating SERP S1 quality through mono-dimensional metrics, and **(c)-(d)** contrasting with multi-dimensional metrics. **Note:** in (c) the usability series values are multiplied by 3 for the illustration.

**Legend for Figures 2 – 4:**

Figures 2 – 4 illustrate the behavior of the framework based on abstract synthetic evaluation data: the relevance themes and attributes are abstract like in Tables 2 – 5. The values are synthetic but quite possible. Appendix I gives the tables that produce the graphs. The graph titles are coded to indicate which type of evaluation is illustrated: the code has positions for each of relevance Overlap, Attributes, and Rank based discount. In each position, a capital letter O/A/R indicates that the corresponding type of discount is applied and a lowercase letter that it is not. For example, “OAr” indicates overlap and attribute discounts applied, but rank-based (= NDCG) turned off.

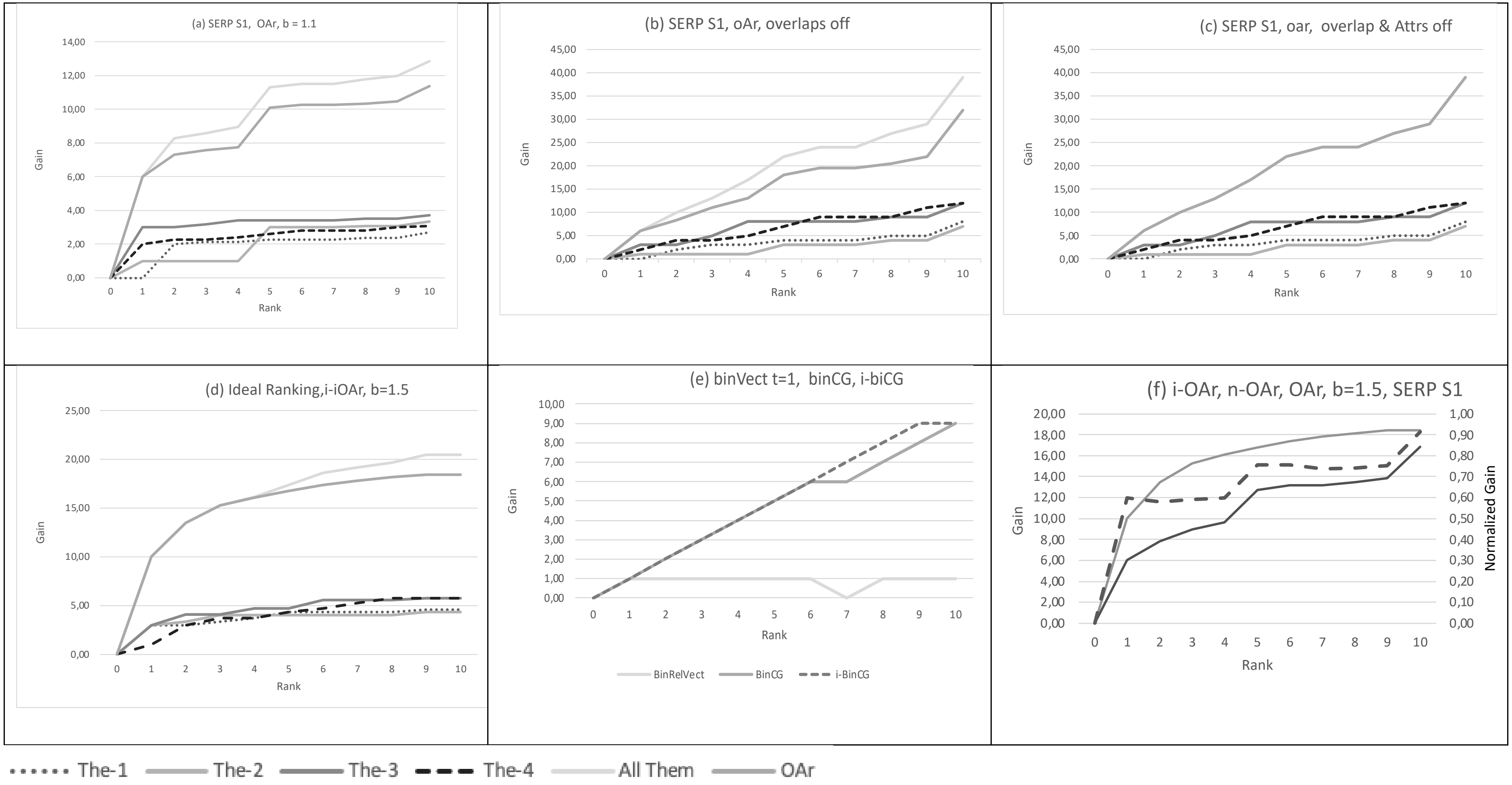

**Figure 2. (a)-(c)** Effectiveness plots with/out overlap discount, attribute downgrading, and **(d)-(f)** various ideal and reference curves, mostly based on SERP S1.

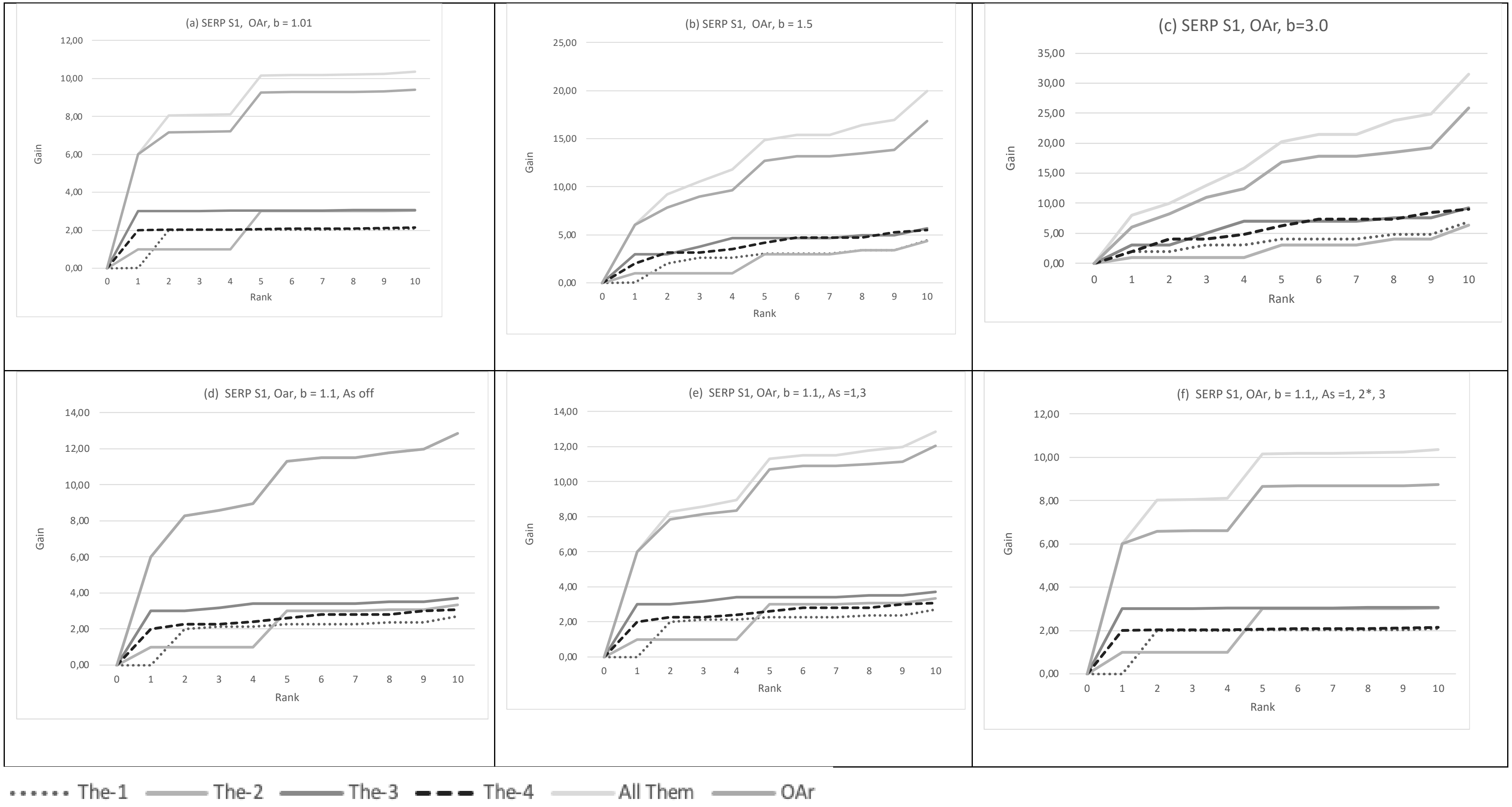

**Figure 3. (a)-(c)** Effect of varying overlap discount, w/o attribute downgrading, and **(d)-(f)** varying attribute values, all based on SERP S1.

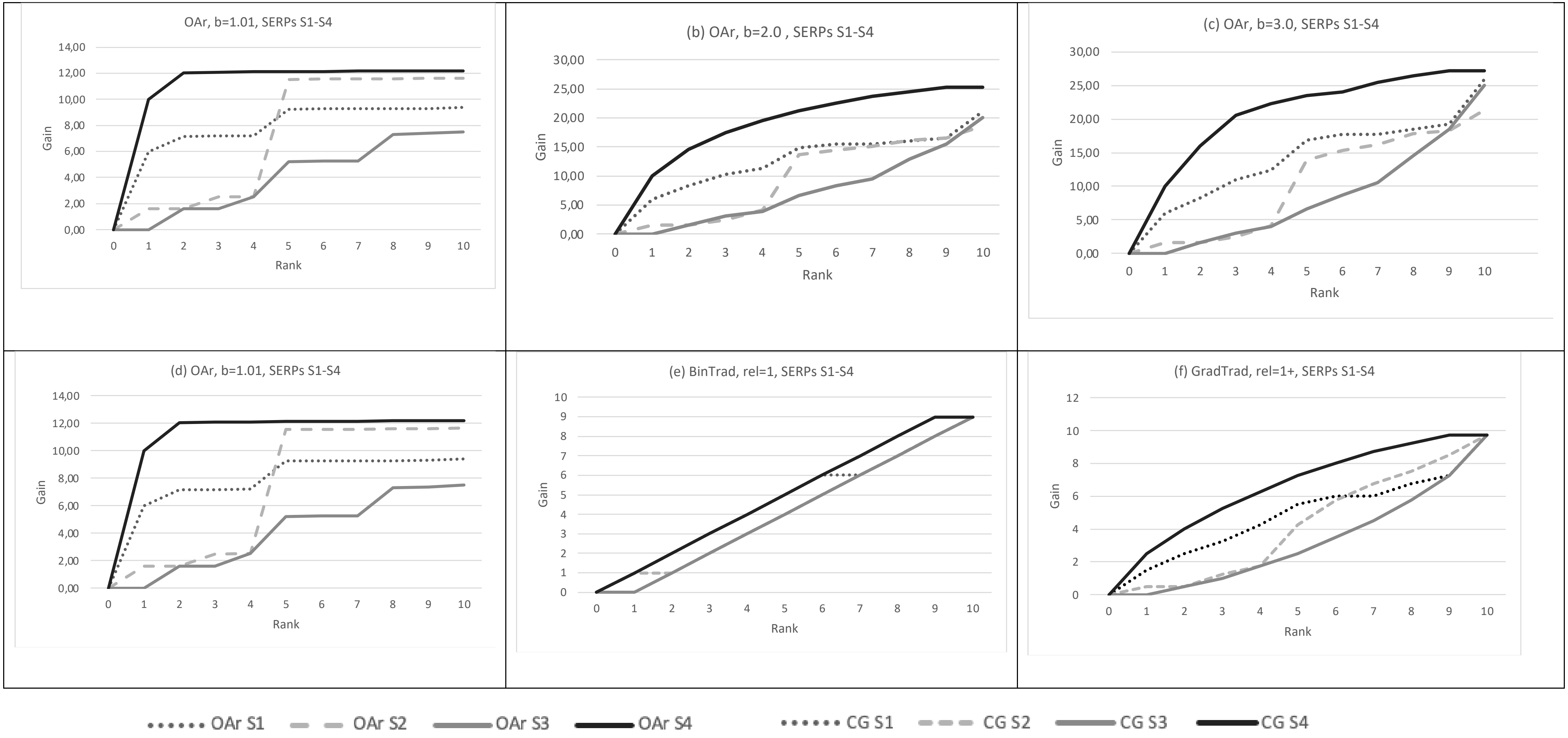

**Figure 4. (a)-(c)** Effect of overlap discount on SERP quality, and **(d)-(f)** contrasting with traditional metrics.

## 5.2. Behavior of the MDCU

*The Behavior and Responsiveness of the Metric*. The validity of MDCU is supported by its behavior as expected under various conditions. Figure 2 (a) – (c) shows the performance of SERP S1 when moderate overlap and attribute discounts are (a) on and (b-c) off. The top curves in each graph are for cross-theme gain and the bottom four for individual themes. Overall we note that the level and shape of the curves change according to which MDCU discounts are in and off. Figure 2 (d) – (f) illustrates the ideal MDCU of S1 in contrast to several reference curves. In (e) with binary gains there is no indication of overlaps for stopping – S1 appears to perform very well. In (f) shows S1 much below the ideal (solid dark curve) and its normalized (dotted) curve between 0.5 – 0.95 of the ideal gain (vertical axis on the right).

Figure 3 (a) – (c) shows the effect of overlap on SERP S1 performance: With heavy overlap (a), the first 1 – 2 documents bring almost all non-overlapping gain per theme and in total while being liberal with overlap (b-c) rewards for lower ranked overlapping results. Themes 3 and 4 seem to perform better when overlap is ignored ($b > 5$). Figure 3 (d) – (f) shows further examples on SERP S1 performance: (d) under moderate overlap and attributes off, (e-f) under heavy vs. liberal overlap and modified attributes (see Appendix).

Figure 4 compares the performance of four SERPs S1 – S4 given in the Appendix under varying conditions. The SERPs contain the same 10 records, but in different orders. They could be results of, e.g., different search engines for the same search task. The top row, (a)-(c), illustrates the effect of overlap discount on SERP scoring. We note that S1 and S2 change orders of goodness depending on how much overlap is given weight The bottom row, (d)-(f), contrast MDCU with traditional metrics. We note that CG based on liberal binary relevance, (e), does not clearly differentiate the SERPs well while the curves based on graded relevance resemble those of MDCU with liberal overlap but differ from the stringent overlap case (d). This is logical.

The examples show that the framework is responsive to the variation of central parameters *b* and attribute factors. The response is also logical. The curves for MDCU turn horizontal for each theme and an entire search result as soon as no further non-overlapping results are expected. This can be much earlier than the last relevant but overlapping results which are credited by the traditional metrics.

The traditional metrics do not suggest where the problem lies, when relevant content is not properly ranked from the task perspective.. The MDCU framework can point out at least which relevance themes or attributes are associated with unsuccessful ranking. The metrics favor search results that maximize gain in a concise result – do not reward for repetition of marginal information. The framework is conceptually broader than the traditional one. It brings new concepts – overlap, attributes, and relevance themes – to IR evaluation, offering a systematic way to bring person and task related factors to IR evaluation.

On the other hand, the proposed framework is general, offering the possibility of deriving traditional evaluation frameworks as special cases. The attributes can be turned off by setting them all at one. Overlap can be ignored by setting the overlap discount parameter *b* high (>5). Mono-dimensional relevance is available by through summing up (perhaps selected) theme relevance scores and setting thresholds for binary/graded relevance.

## 5.3. Building the MDCU framework

*Building the Test Collection – Search Tasks in Context.* The proposed framework supports designing task-situation-person based search tasks. The relevance themes are a natural way of representing dimensions of the person's larger task around the search task. Analyses of TBII (see Section 2.2) in real-life contexts help here. The themes or entire search tasks may be classified and selected or excluded in a specific experiment to represent the person's situation in, and knowledge of, the larger task. Following the Title – Description – Narrative (TDN) structure of a search task description, the narrative can contain vocabularies per theme – supporting algorithmic (assistance in) overlap assessment when building the relevance corpus. TBII studies and search engine logs are helpful in specifying the attributes. It seems reasonable to expect that the MDCU framework is the more appropriate the more complex the search task is.

*Building the Test Collection – Relevance Corpus.* The relevance corpus may be derived from a traditional relevance corpus by extended analysis. All at least marginally relevant documents would be reanalyzed for task-situation-person related aspects. The scoring of the relevance themes could be similar to SOTA in test-collection construction -- now only more reliable due to the structure provided by the themes -- either through human intellectual judgment, crowdsourcing or statistical/algorithmic assistance. The scoring of usability attributes can often happen algorithmically (the language, difficulty, bias, difficulty, …) but require in some cases specification of the data source and its analysis functions (e.g., social media or recommender systems data).

*Limitations -- Validity.* Traditional evaluation is constrained due to naïve assumptions and because of not bringing in new concepts. The assumptions within MDCU are bold, especially regarding discounting by accumulated gain. However, the framework allows what-if experiments where the overlap assumptions are varied. Its scoring behavior also seems logical in response to variations in the input. Moreover, the assumptions can be empirically tested for validity in various circumstances. After all – which is more valid an assumption: (a) "all documents are equal in cumulating gains no matter in what their position in the result" or (b) "the more gain you have collected, the less there remains novel gains to collect"? The MDCU framework therefore seems valid, if the additional effort can be afforded and the evaluator gets the input numbers right.

*Limitations -- Cost.* The test collections required by MDCU are clearly more complex than the traditional ones. Their construction requires more effort. In return the collections allow more exciting research as output. We have no cost estimates nor a convincing large empirical case to present. Therefore the word 'blueprint' in the title. However, IR is also a multi-billion-dollar business, so the relative cost is marginal. Also graded relevance assessments were initially opposed for cost reasons but are part of the SOTA in IR evaluation for years now. It is possible to support theme relevance scoring through measuring document (passage) distances from the search task theme specifications (in the N component). Likewise attribute scoring could be assisted through machine learning on the user profiles of a relevant context. Meanwhile overlap handling could be refined through empirical studies on overlaps of various types of documents.

*Limitations -- Not needed*: One might argue that there is no need for MDCU , because the multidimensional relevance case can be viewed as a fusion problem of theme-based document steams. Indeed, this is one way of developing the search and rank algorithm but has no direct bearing on the evaluation.

## 5.4. Experimentation with the MDCU framework

*Support to Systematic Experimentation.* Serving the goal of helping the person performing a task requires bringing some concepts representing task, situation, and person characteristics into IR evaluation. Otherwise we have fairy tale. The MDCU framework opens IR evaluation for new concepts, their operationalizations, and fosters systematic experimentation with new questions. Here are some examples:

- [Overlap estimation / discount:] How significant is the overlap problem in various types of collections? How much overlap is there and how serious is the bias in evaluation results under various conditions?
- [Attribute discounts:] Can the text classification methods for labeling documents as difficult, biased, or sentimental be effectively used for demoting less usable documents in search results?
- [Identifying relevance themes:] How does the extent of searcher's knowledge within/across themes affect searching / results? For learning in a session, which themes/concepts is the searcher exposed to, when searching on some other?
- [Classification of relevance themes:] Which types of themes or search keys work best as handles to other themes? What kind of themes are easiest to rank high / accumulate early?

## 5.5. MDCU with other areas of IR

*Support to Integration with Other Research Areas.* Several research areas within or close to the broad area of IR could benefit from search engines which are effective by the MDCU criteria. For example

- [Question Answering Systems:] There are mutual benefits in developing methods for finding concise answers to queries: less NLP needed on the QA side, and empirical support in deciding relevance themes and attributes for QA type of IR tasks.

- [Recommender Systems:] Data on how recommender systems structure their users' interactions can be used in deciding about relevance themes and attributes, and typical attribute values for RecSys type of IR tasks; methods for producing concise summaries from textual product reviews.
- [Searching as Learning:] Suppose that we have a search task on three concepts A, B, C with their expressions {…, $a_i$, …}, {…, $b_j$, …}, {…, $c_k$, …}, respectively, available. Suppose further that the searcher only knows weakly about the concept of A, some expressions. When the searcher's interactions are observed, analyze to which concepts he/she was exposed in the documents seen, and whether they were later used / remembered.

There are more constrained and more structured information environments, like e-commerce, where the relevance themes are more obvious and persistent than in general Web-IR. We leave this for later.

## 6. Conclusion

We have argued that IR evaluation is not serving IR research and development well, because of the assumptions it makes regarding relevance, document content overlap,, cumulation of gain in search results, and the role of document usability attributes. If all search engines perform well in SOTA test collection-based evaluation, then the search tasks are too easy. The evaluation does not test retrieval methods for being helpful for the searcher in her/his motivating larger task. More demanding goals need to be set. They require a new evaluation framework. We propose

> Giving concise non-overlapping answers that are helpful for a person or software application in a situation while performing a task.

as the goal of IR and

> Measuring how well a search engine achieves this IR goal by retrieving and ranking documents, components, or information.

as the goal of IR evaluation. This requires that task – situation – searcher specific factors are made explicit in IR. We therefore propose a new evaluation framework which uses multidimensional and graded theme relevance and attributes to represent these factors. Following Newell (1993), we have presented the evaluation task, the definition of some of its algorithms, and their justification. The framework offers novel evaluation metrics, called MDCU metrics (for multidimensional cumulative gain), which allow control of the task, situation, and searcher factors, and document overlap, in evaluation. The algorithms are defined formally and justified by examples indicating their behavior as desired in comparison to traditional IR evaluation. The limitations were discussed: the costs of building the required test collections clearly exceed the costs of building traditional ones of comparable size. In return the framework fosters more demanding and valid IR evaluation.

The contributions of the paper are:

- Proposal on integrating task – situation – searcher attributes as factors into IR evaluation.
- Proposal on avoiding naive assumptions regarding document relevance and overlap.
- Proposing a general approach: traditional IR evaluations as special cases; service to other IR subdomains where search engines serve as components.

## Appendix: Document lists

**SERP S1**

| I:oAr | Theme Relevances | | | | Usability Attributes | | |
|---|---|---|---|---|---|---|---|
| **Doc id#** | **Theme1** | **Theme2** | **Theme3** | **Theme4** | **Attr1** | **Attr2** | **Attr3** |
| **d1** | 0,00 | 1,00 | 3,00 | 2,00 | 1,00 | 1,00 | 1,00 |
| **d2** | 2,00 | 0,00 | 0,00 | 2,00 | 0,90 | 0,70 | 0,90 |
| **d3** | 1,00 | 0,00 | 2,00 | 0,00 | 1,00 | 0,90 | 1,00 |
| **d4** | 0,00 | 0,00 | 3,00 | 1,00 | 0,80 | 0,90 | 0,70 |
| **d5** | 1,00 | 2,00 | 0,00 | 2,00 | 1,00 | 1,00 | 1,00 |
| **d6** | 0,00 | 0,00 | 0,00 | 2,00 | 1,00 | 0,80 | 1,00 |
| **d7** | 0,00 | 0,00 | 0,00 | 0,00 | 0,00 | 0,00 | 0,00 |
| **d8** | 1,00 | 1,00 | 1,00 | 0,00 | 0,30 | 1,00 | 1,00 |
| **d9** | 0,00 | 0,00 | 0,00 | 2,00 | 0,90 | 0,90 | 0,90 |
| **d10** | 3,00 | 3,00 | 3,00 | 1,00 | 1,00 | 1,00 | 1,00 |

**SERP S1 with mono-dimensional relevance**

| Doc id# | Average relevance | P @ R (>1) | P @ R (>2) | Total relevance | Total CG | Average CG | |
|---|---|---|---|---|---|---|---|
| d1 | 1,50 | 1,00 | 0 | 6 | 6 | 1,5 | |
| d2 | 1,00 | 1,00 | 0 | 4 | 10 | 2,5 | |
| d3 | 0,75 | 0,67 | 0 | 3 | 13 | 3,25 | |
| d4 | 1,00 | 0,75 | 0 | 4 | 17 | 4,25 | |
| d5 | 1,25 | 0,80 | 0 | 5 | 22 | 5,5 | |
| d6 | 0,50 | 0,67 | 0 | 2 | 24 | 6 | |
| d7 | 0,00 | 0,57 | 0 | 0 | 24 | 6 | |
| d8 | 0,75 | 0,50 | 0 | 3 | 27 | 6,75 | |
| d9 | 0,50 | 0,44 | 0 | 2 | 29 | 7,25 | |
| d10 | 2,50 | 0,50 | 0,1 | 10 | 39 | 9,75 | |

**Ideal Ranking**

| :oAr | **Theme Relevances** | | | | **Usability Attributes** | | |
|---|---|---|---|---|---|---|---|
| **Doc id#** | **Theme1** | **Theme2** | **Theme3** | **Theme4** | **Attr1** | **Attr2** | **Attr3** |
| **ir1 d10** | 3,000 | 3,000 | 3,000 | 1,000 | 1,00 | 1,00 | 1,00 |
| **ir2 d1** | 0,000 | 0,369 | 1,107 | 2,000 | 1,00 | 1,00 | 1,00 |
| **ir3 d5** | 0,369 | 0,668 | 0,000 | 0,738 | 1,00 | 1,00 | 1,00 |
| **ir4 d3** | 0,334 | 0,000 | 0,574 | 0,000 | 1,00 | 0,90 | 1,00 |
| **ir5 d2** | 0,619 | 0,000 | 0,000 | 0,615 | 0,90 | 0,70 | 0,90 |
| **ir5 d4** | 0,000 | 0,000 | 0,861 | 0,369 | 0,80 | 0,90 | 0,70 |
| **ir4 d6** | 0,000 | 0,000 | 0,000 | 0,522 | 1,00 | 0,80 | 1,00 |
| **ir5 d9** | 0,000 | 0,000 | 0,000 | 0,489 | 0,90 | 0,90 | 0,90 |
| **ir5 d8** | 0,277 | 0,291 | 0,237 | 0,000 | 0,30 | 1,00 | 1,00 |
| **ir5 d7** | 0,000 | 0,000 | 0,000 | 0,000 | 0,00 | 0,00 | 0,00 |

**S1 but Modified Attribute scoring**

| Usability Attributes Figure 2 e | Usability Attributes Figure 2 f | Usability Attributes not used |
|---|---|---|

| Attr1 | Attr2 | Attr3 | Attr1 | Attr2 | Attr3 | Attr1 | Attr2 | Attr3 |
|---|---|---|---|---|---|---|---|---|
| 1,00 | 1,00 | 1,00 | 1 | 1,00 | 1 | 1 | 1 | 1 |
| 0,90 | 0,35 | 0,90 | 0,9 | 1,00 | 0,9 | 1 | 1 | 1 |
| 1,00 | 0,90 | 1,00 | 1 | 1,00 | 1 | 1 | 1 | 1 |
| 0,80 | 0,45 | 0,70 | 0,8 | 1,00 | 0,7 | 1 | 1 | 1 |
| 1,00 | 1,00 | 1,00 | 1 | 1,00 | 1 | 1 | 1 | 1 |
| 1,00 | 0,40 | 1,00 | 1 | 1,00 | 1 | 1 | 1 | 1 |
| 0,00 | 0,00 | 0,00 | 0 | 1,00 | 0 | 1 | 1 | 1 |
| 0,30 | 0,50 | 1,00 | 0,3 | 1,00 | 1 | 1 | 1 | 1 |
| 0,90 | 0,90 | 0,90 | 0,9 | 1,00 | 0,9 | 1 | 1 | 1 |
| 1,00 | 0,50 | 1,00 | 1 | 1,00 | 1 | 1 | 1 | 1 |
| | | | | | | | | |

**SERP S2 Halves swapped**

| II:OAr | Theme Relevances with Overlap Discounts | | | | Usability Attributes | | |
|---|---|---|---|---|---|---|---|
| **Doc id#** | Theme1 | Theme2 | Theme3 | Theme4 | Attr1 | Attr2 | Attr3 |
| **Doc6** | 0,000 | 0,000 | 0,000 | 2,000 | 1 | 0,80 | 1 |
| **Doc7** | 0,000 | 0,000 | 0,000 | 0,000 | 0 | 0,00 | 0 |
| **Doc8** | 1,000 | 1,000 | 1,000 | 0,000 | 0,3 | 1,00 | 1 |
| **Doc9** | 0,000 | 0,000 | 0,000 | 2,000 | 0,9 | 1,00 | 0,9 |
| **Doc10** | 3,000 | 3,000 | 3,000 | 0,792 | 1 | 1,00 | 1 |
| **Doc1** | 0,000 | 0,792 | 2,377 | 1,402 | 1 | 0,30 | 1 |
| **Doc2** | 1,585 | 0,000 | 0,000 | 1,205 | 0,9 | 0,40 | 0,9 |
| **Doc3** | 0,639 | 0,000 | 1,186 | 0,000 | 1 | 0,90 | 1 |
| **Doc4** | 0,000 | 0,000 | 1,629 | 0,549 | 0,8 | 0,35 | 0,7 |
| **Doc5** | 0,601 | 1,402 | 0,000 | 1,060 | 1 | 1,00 | 1 |

**SERP S3 Worst**

| II:OAr | Theme Relevances with Overlap Discounts | | | | Usability Attributes | | |
|---|---|---|---|---|---|---|---|
| **Doc id#** | Theme1 | Theme2 | Theme3 | Theme4 | Attr1 | Attr2 | Attr3 |
| **Doc7** | 0,00 | 0,0 | 0,000 | 0,00 | 0,00 | 0,00 | 0,00 |
| **Doc6** | 0,00 | 0,00 | 0,00 | 2,00 | 1,00 | 0,80 | 1,00 |
| **Doc9** | 0,00 | 0,00 | 0,00 | 2,00 | 0,90 | 0,90 | 0,90 |
| **Doc8** | 1,00 | 1,00 | 1,00 | 0,00 | 0,30 | 1,00 | 1,00 |
| **Doc3** | 1,00 | 0,00 | 2,00 | 0,00 | 1,00 | 0,90 | 1,00 |
| **Doc2** | 2,00 | 0,00 | 0,00 | 1,58 | 0,90 | 0,70 | 0,90 |
| **Doc4** | 0,00 | 0,00 | 3,00 | 0,64 | 0,80 | 0,90 | 0,70 |
| **Doc5** | 0,79 | 2,00 | 0,00 | 1,20 | 1,00 | 1,00 | 1,00 |
| **Doc1** | 0,00 | 1,00 | 1,84 | 1,10 | 1,00 | 1,00 | 1,00 |
| **Doc10** | 2,10 | 2,38 | 1,60 | 0,51 | 1,00 | 1,00 | 1,00 |

**SERP S4 Best**

| II:OAr | **Theme Relevances with Overlap Discounts** | | | | **Usability Attributes** | | |
|---|---|---|---|---|---|---|---|
| **Doc id#** | Theme1 | Theme2 | Theme3 | Theme4 | Attr1 | Attr2 | Attr3 |
| **Doc10** | 3,00 | 3,00 | 3,00 | 1,00 | 1,00 | 1,00 | 1,00 |
| **Doc1** | 0,00 | 1,00 | 3,00 | 2,00 | 1,00 | 1,00 | 1,00 |
| **Doc5** | 1,00 | 1,58 | 0,00 | 2,00 | 1,00 | 1,00 | 1,00 |
| **Doc2** | 1,58 | 0,00 | 0,00 | 1,37 | 0,90 | 0,70 | 0,90 |
| **Doc4** | 0,00 | 0,00 | 1,84 | 0,59 | 0,80 | 0,90 | 0,70 |
| **Doc8** | 0,64 | 0,64 | 0,53 | 0,00 | 0,30 | 1,00 | 1,00 |
| **Doc3** | 0,60 | 0,00 | 1,03 | 0,00 | 1,00 | 0,90 | 1,00 |
| **Doc6** | 0,00 | 0,00 | 0,00 | 1,13 | 1,00 | 0,80 | 1,00 |
| **Doc9** | 0,00 | 0,00 | 0,00 | 1,05 | 0,90 | 0,90 | 0,90 |
| **Doc7** | 0,00 | 0,00 | 0,00 | 0,00 | 0,00 | 0,00 | 0,00 |